\documentclass{iau}
\usepackage{graphicx}

\title[Algol-type systems V1241 Tau and GQ Dra] 
{Asteroseismic Investigation of two Algol-type systems V1241 Tau and GQ Dra}

\author[Ula\c{s} et al.]   
{Burak Ula\c{s}$^1$, Ceren Ulusoy$^2$, Kosmas Gazeas$^3$, Naci Erkan$^4$, Alexios Liakos$^5$}

\affiliation{$^1$\.{I}zmir Turk College Planetarium, 8019/21 sok., No: 22, \.{I}zmir, Turkey \\ email: {\tt bulash@gmail.com} \\[\affilskip]
$^2$College of Graduate Studies, University of South Africa, PO Box 392, UNISA 0003, Pretoria, South Africa \\email: {\tt cerenuastro@gmail.com} \\[\affilskip]
$^3$Department of Astrophysics, Astronomy and Mechanics, National and Kapodistrian University of Athens, GR-157 84, Zografos, Athens, Greece  \\email: {\tt kgaze@phys.uoa.gr} \\[\affilskip]
$^4$Department of Physics, Faculty of Arts and Sciences, \c{C}anakkale Onsekiz Mart University, Terzioglu Campus, TR-17100, \c{C}anakkale, Turkey \\email: {\tt nacierkan@comu.edu.tr} \\[\affilskip]
$^5$Institute for Astronomy \& Astrophysics, Space Applications \& Remote Sensing, National Observatory of Athens, I.Metaxa \& Vas. Pavlou St., GR-15236, Palaia Penteli, Greece \\email: {\tt alliakos@phys.uoa.gr}}

\pubyear{2008}
\volume{xxx}  
\pagerange{119--126}
\jname{Title of your IAU Symposium}
\editors{A.C. Editor, B.D. Editor \& C.E. Editor, eds.}
\begin{document}

\maketitle

\begin{abstract}
We present new photometric observations of eclipsing binary systems V1241 Tau and GQ Dra. We use the following methodology: Initially, WD code is applied to the light curves, in order to determine the photometric elements of the systems. Then the residuals are analysed using Fourier Transformation techniques. The results show that one frequency can be barely attributed to the residual light variation of V1241 Tau, while there is no evidence of pulsation on the light curve of GQ Dra.
\end{abstract}

\firstsection 
\section{Introduction}

V1241 Tau was first observed by Henrietta Leavitt (\cite{yan12}). Although \cite{rod00} defended a $\delta$ Sct type variation on the light curve of the system, \cite{are04} mentioned that no trace of pulsation can be seen. 
The light variation of GQ Dra was first determined by Hipparcos (\cite{esa97}). 

\section{Observations and Solution of the Light Curves}

\begin{figure}
\begin{center}
\includegraphics{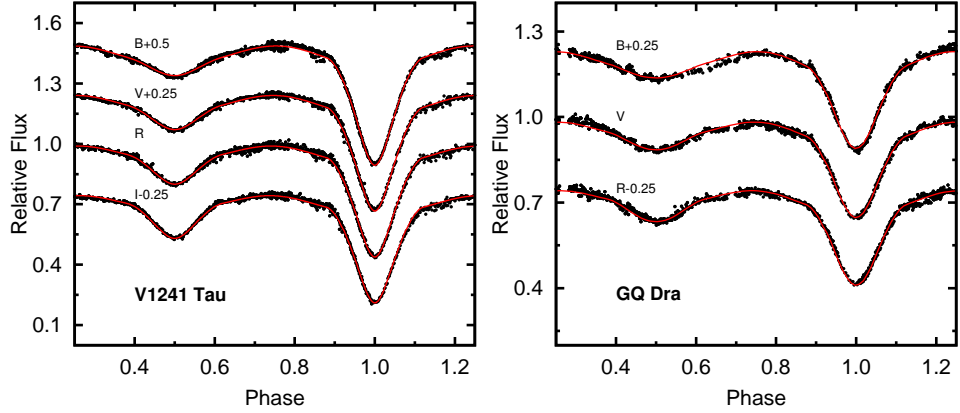}  
\caption{Observed (points) and theoretical (lines) light curves of the systems. Some curves are shifted in flux axis for the sake of clear visibility.}
\label{fig1}
\end{center}
\end{figure}

BVRI light curves of V1241 Tau were obtained with 0.4-m telescope of the University of Athens Observatory in November 2012.  
GQ Dra was observed with the 1.22-m telescope of the Onsekiz Mart University Observatory 
in 7 nights between March-April 2013. 
Light curves of both systems were analysed using PHOEBE (Pr\v{s}a and Zwitter 2005) software. 
Results show that the system V1241 Tau has a semi-detached configuration where the inclination is about 81$^o$.5 and the mass ratio is 0.44. The hotter and cooler components has the temperature values of 7500 K and 4906 K, respectively. We concluded that 95 percent of the light in $V$ filter comes from the primary component. Our solution of semi-detached binary GQ Dra is the first light curve solution in the literature. Mass ratio of the system is found to be 0.25 and the orbital inclination was calculated as 75$^o$.3 during the solution. The temperature of the secondary derived as 5050 K where the primary's value was fixed to 8750 K. The agreement between results and the observations is drawn in Fig.~\ref{fig1}.

\section{Search for Pulsations}

We tried to represent the residuals with a periodic variation using Fourier Transformation techniques. In the case of V1241 Tau, our research has resulted in one frequency (f=2.13 c/d) which can be attributed to a change in the light curve. However, it is worth to emphasize that this kind of variation may be assigned to other reasons as well, such as observational effects and atmospheric conditions. The same method was applied to the residuals yielded from the binary solution of GQ Dra which showed no remarkable oscillation-like variation.

\section{Conclusions}

The main result regarding the pulsational behaviour of the systems is that there is no trace of pulsations in neither of them. The residual light curve of the system V1241 Tau can be represented by a periodic variation, however, it is not satisfactorily enough to mention any physical (low amplitude) oscillation and it may be assigned to other reasons which mentioned in the previous section.

~\\
{\bf{Acknowledgements}}\\
\small{CU sincerely thanks the South African National Research Foundation (NRF) for the award of NRF MULTI-WAVELENGTH ASTRONOMY RESEARCH PROGRAMME (MWGR), Grant No: 86563 to Prof LL Leeuw at UNISA, Reference: MWA1203150687.}

\end{document}